\documentstyle[psfig,12pt,aas2pp4]{article}
\begin{document}
\title{A persistent $\sim$1 Hz quasi-periodic oscillation in
the dipping low-mass X-ray binary 4U\,1323--62}

\author{Peter G. Jonker\altaffilmark{1}, Michiel van der
Klis\altaffilmark{1}, Rudy Wijnands\altaffilmark{1}}
\altaffiltext{1}{Astronomical Institute ``Anton Pannekoek'',
University of Amsterdam, and Center for High-Energy Astrophysics,
Kruislaan 403, 1098 SJ Amsterdam; peterj@astro.uva.nl,
michiel@astro.uva.nl, rudy@astro.uva.nl}
\subjectheadings{accretion, accretion disks --- stars: individual
(4U1323-62) --- stars: neutron --- X-rays: stars}
\begin{abstract}
We have discovered a $\sim$1 Hz quasi-periodic oscillation (QPO) in
the persistent-emission, the dips, and the type I X-ray bursts of the
low-mass X-ray binary 4U\,1323--62. The rms amplitude of the QPO is
approximately 9\%, only weakly depending on photon energy. The
amplitude is consistent with being constant throughout the
persistent-emission, the dips and the bursts in all but one
observation, where it is much weaker during one dip. These properties
suggest that we have observed a new type of QPO, which is caused by
quasi-periodic obscuration of the central X-ray source by a structure
in the accretion disk. This can only occur when the binary inclination
is high, consistent with the fact that 4U\,1323--62 is a dipping
source. The quasi-periodic obscuration could take place by partial
covering of an extended central X-ray source by a near-opaque medium,
or by covering of a point source by a medium of suitable
characteristics to produce the relatively energy-independent
oscillations.
\end{abstract}

\section{Introduction}
The low-mass X-ray binary (LMXB) 4U\,1323--62 shows both type I X-ray
bursts and periodic dips. The bursts and dips were discovered with the
EXOSAT satellite (van~der~Klis et al. 1984, 1985a,b). The periodic
dips in the light curves of dipping LMXBs are thought to be caused by
periodic obscuration of the central source by a structure formed in an
interplay between the gas stream from the companion and the accretion
disk (White~\&~Swank 1982; White, Nagase, \& Parmar 1995); the
period of the dips is the orbital period.
\newline
\indent In 4U\,1323--62, the dip period is 2.93 hours and the dips
last $\sim$50 minutes (van der Klis et al. 1985a,b; Parmar et
al. 1989). From the burst properties and peak fluxes, and from the
absence of eclipses the distance of the source is inferred to be
10--20 kpc, and the inclination $\le$ 80$^\circ$. The energy spectrum
fits an absorbed power law with energy spectral index $\alpha
= 0.53\pm0.07$, and $\rm{N_H} = (4.0\pm0.3) \rm{x} 10^{22}$ atoms
$\rm{cm}^{-2}$.
\newline
\indent In non-dipping LMXBs, numerous QPO phenomena have been
reported, at frequencies of several Hertz to $>$ 1200 Hz (van der Klis
1995, 1998 for reviews). So far observations of any $<$ 100 Hz QPO
phenomena in dipping LMXBs have been lacking (kHz QPOs were recently
reported in 4U\,1916--05 by Barret et al. 1997).  In this Letter we
report the discovery of $\sim$1 Hz QPOs in the dipper
4U\,1323--62. The frequency as well as the amplitude of the QPO is
approximately constant thoughout the bursts, dips, and
persistent-emission. This is the first QPO reported to persist both
through type I X-ray bursts and the persistent-emission, and the first
$<100$ Hz frequency QPO in a dipper.

\section{Observations and analysis}
4U\,1323--62 was observed five times with the proportional counter
array (PCA; Jahoda et al. 1996) on board the RXTE (Rossi X-ray Timing
Explorer; Bradt, Rothschild \& Swank 1993) satellite on 1997 April 25,
26, and 27 (see Table \ref{obs_log}). The total amount of good data
was 80 ksec. During each observation one or two bursts and one or two
dips were observed, resulting in a total of seven type I X-ray bursts
and eight dips (some of which were only partially observed). Two
bursts in observation 3 showed a secondary burst $\sim$500 s after the
primary one. These last approximately 30 s, rise quikly up to count
rates of 370 c/s (2--60 keV; this is $\sim$21\% of the primary burst
peak count rate) and 650 c/s ($\sim$37\%) and decrease slowly. A burst
in observation 2 took place during a dip. The dips lasted $\sim$60
minutes, slightly longer than those reported by Parmar et
al. (1989). The mean background-subtracted persistent 2--60 keV count
rate changed little between observations (see Table
\ref{obs_log}). In observation 4, during 40\% of the time only 4 of
the 5 PCA detectors were active. All reported count rates are for 5
detectors.
\newline
\indent
Data were obtained over an energy range of 2--60 keV with 16
s time resolution in 129 energy channels, and simultaneously with a
time resolution of 1 $\mu$s in 255 energy channels. 
\newline
\indent In Figure \ref{figlc} (top panel), we show part of the light
curve of observation 1, showing a dip, a burst, and the
persistent-emission. The dip starts $\sim$600 s after the beginning of
the observation and lasts untill $\sim$4300 s. We also show the
corresponding hardness curve (lower panel); the hardness is defined as
the 16-s averaged ratio of the count rate in the 9.7--16 keV energy
band to that in the 2--7.9 keV band. During the dips the hardness
ratio is higher than in the persistent emission, and at the burst
start the hardness ratio decreases.
\newline
\indent Using the light and hardness curves, we divided the data into
three different categories; the dips, the bursts and the persistent
emission. This led to the definition of time intervals during which
the source is in one of these three states. When analyzing the data
the same time intervals were applied in each energy band.  We defined
dip ingress by the decrease in count rate and simultaneous increase in
hardness evident in Figure \ref{figlc}; dip egress is the inverse. In
between dip ingress and egress, the count rate and hardness behavior
is erratic during $\sim$60 minutes. Sometimes during a dip the
persistent-emission count rate and hardness levels are briefly
reached. We still considered those data to be part of the dip. The
onset of a burst is characterized by a steep rise in the count rate
simultaneously with a decrease in the hardness ratio. We determined
the burst decay time by fitting an exponential to the (2--60 keV)
decay at a time resolution of 1/8th second. The end of the burst was
taken as three times the e-folding time after the onset. One burst
decay time was $60\pm2$ s, five were consistent with 82 s and one was
$92\pm3$ s. In our power spectral analysis of the bursts concerning
the $\sim$1 Hz QPO, we only used the burst decay interval; in our
search for $>100$ Hz burst oscillations we also used the burst rises.
\newline
\indent We calculated power spectra separately for the
persistent-emission and for the dips using 64-s segments with a time
resolution of 1 ms in the 2--60 keV band, as well as in four energy
bands (2--5.0, 5.0--6.4, 6.4--8.6, 8.6--13.0 keV). We applied a
similar analysis to the bursts, but using 16-s segments enabling us to
obtain 5 power spectra for most of the bursts, and reducing the low
frequency component in the power spectra due to the burst profile.
\newline
\indent In each energy band all 1/64--512 Hz power spectra
corresponding to dips (320 in number), or persistent emission (881)
were averaged. The power spectra were fitted with a fit function
consisting of a Lorentzian (the QPO), an exponentially cut-off power
law, and a constant to represent the Poisson noise. In case of the
(1/16--512 Hz) power spectra calculated from the 16-s segments (32)
obtained during bursts, an extra Lorentzian centered on $\sim$0 Hz was
used in order to account for the power spectral component due to the
burst profile. The errors on the fit parameters were determined using
$\Delta\chi^2 = 1.0$ (1$\sigma$ single parameter), and upper limits by
using $\Delta\chi^2 = 2.71$, corresponding to a 95\% confidence level.

\section{Results}
In the persistent-emission power spectra, we discovered a very
significant (31 $\sigma$) $\sim$1 Hz QPO (Fig. \ref{fig1Hz}). Its
frequency was $0.87\pm0.01$ Hz in observation 1, 2, and 3, and
$0.77\pm0.01$ Hz in observation 4 and 5.  This shift of $\sim$0.1 Hz
between observations is not correlated to the changes in the
persistent count rate level (Table \ref{obs_log}). The QPO
was detected during the persistent emission as well as during the dips
and the bursts. It can be directly observed in the light curves of the
bursts (see Fig. \ref{light_b_1Hz}). During a burst the amplitude of
the QPO increases by a factor of $\sim$10 to keep the fractional
amplitude approximately constant (see below). The Poisson counting
noise prevented us to directly see the QPO in the non-burst parts of
the lightcurve.
\newline
\indent Except during the dip in observation 4 when the fractional rms
amplitude was only $2.2\pm0.2$\% and the FWHM only $0.07\pm0.02$ Hz,
the frequency, 2--60 keV fractional rms amplitude, and FWHM of the QPO
are consistent with being the same for the dips and the
persistent-emission in each observation at values of 0.77 or 0.87 Hz,
9\%, and 0.25 Hz, respectively. To improve the signal to noise, we
averaged the power spectra corresponding to the persistent emission
and the dips across all observations in our further analysis. The FWHM
and frequency of the QPO in all the persistent-emission and dip data
combined is $0.25\pm0.01$ Hz and $0.84\pm0.02$ Hz, respectively; in
the individual energy bands we found numbers consistent with this. In
determining the fractional rms amplitudes reported below, we fixed
FWHM and frequency to these values.  The 2--60 keV fractional rms
amplitudes in all observations combined have values of $9.1\pm0.1$\%,
$8.9\pm 0.4$\%, and $10.9\pm0.5$\% for persistent emission, dips, and
bursts, respectively. While the persistent-emission and the dip rms
amplitudes are identical within the errors, the rms amplitude of the
QPO during the bursts is slightly higher. Systematic errors in the rms
normalization due to the trend in the count rate in the 16-s data
segments caused by the burst profile, and due to the interaction in
the fit procedure with the extra Lorenzian component can probably
account for this small difference. We estimate their influence to be
of the order of the discrepancy, since with 64-s burst power spectra,
where both of these effects are more prominent, we obtained an rms
amplitude of $11.9^{+3.4}_{-0.9}$\%.
\newline
\indent The rms amplitude of the QPO only weakly depends on energy
(Table \ref{fit_table}); in the persistent-emission it is
consistent with a small increase towards higher energies. The
fractional rms amplitudes during the dips, and the bursts are
consistent with this small increase. No time delay was found in the
QPO (0.7--0.9 Hz) between the 2.8--7.5 keV and 7.5--60 keV energy
bands, with a 95\% confidence upper limit of 14.5 ms for a soft lag
($1.8 \rm{x} 10^{-2}$) times the QPO cycle) and an upper limit of 5.4
ms ($6.8 \rm{x} 10^{-3}$ times the QPO cycle) for a hard lag.
\newline
\indent We seached for kHz QPOs, but none were found with upper limits
of 8--10\% for a fixed FWHM of 25 Hz over a frequency range of
100--1000 Hz in all persistent-emission and dip data combined. These
upper limits do not exclude the presence of kHz QPOs since many other
sources have kHz QPO fractional rms amplitudes below this (van der
Klis 1998).
\newline
\indent We set upper limits on any band-limited noise component by
adding an exponentially cut-off power law to the fit function.  Fixing
the power law index to values of 0--1, we derived upper limits of
4.2\%--6.4\%.
\newline
\indent We searched the primary and secondary burst data for high
frequency burst oscillations in various ways. In order to increase the
sensitivity we averaged the burst rise power spectra of the different
bursts using different energy bands (see also Miller 1998). No high
frequency burst oscillations were found with upper limits from 0.5-s
power spectra varying between 24\% at the top and 45\% near the end of
the burst.
\newline
\indent For reference we fitted the persistent-emission energy
spectrum of part of observation 2 with a model consisting of an
absorbed power law and a gaussian line at $6.5 \pm 0.2$ keV with a
FWHM of $\sim$1 keV. The fit was good, with a $\chi^2$ per degree of
freedom of 36/40. The photon index of the power law was $1.75 \pm
0.02$, and the hydrogen column density is $(4.0 \pm 0.3) \rm{x}
10^{22}$ atoms $\rm{cm^{-2}}$. This spectrum implies a 2--25 keV flux
of $1.5 \rm{x} 10^{-10} \rm{erg} \,\rm{s}^{-1}$. Using a distance of
10 kpc this results in a 2--25 keV luminosity of $1.8\rm{x}10^{36}$
erg s$^{-1}$. The spectrum is similar to that previously found by
Parmar et al. (1989), although our photon index is somewhat steeper.

\section{Discussion}
We have discovered a QPO with a frequency of approximately 1 Hz. 
This QPO shows a unique combination of properties;
\newline
\indent 1) it is observed during the bursts, dips and
persistent emission with the same rms amplitude (except during the dip
in observation 4);
\newline
\indent 2) the rms amplitude of the QPO depends only weakly on photon
energy;
\newline
\indent 3) no band-limited noise component was found with upper limits
of $\sim$5\%.
\newline
\indent In all types of LMXBs, QPOs are known to occur with
frequencies ranging from 0.01 to 1200 Hz (see van der Klis 1995, 1998
for reviews). Low frequency (0.01--10 Hz) QPOs are known to occur in a
number of black hole candidates and atoll sources (see van der Klis
1995 for a review). Unlike the QPO we found in 4U\,1323--62, the
fractional amplitudes of low frequency QPO in these sources depend
strongly on photon energy, and the QPOs are found to be superimposed
on a strong band-limited noise component (Wijnands \& van der Klis
1998). In the Z sources (the most luminous neutron star LMXBs) low
frequency (5--20 Hz) QPOs are also well known (van der Klis 1995), the
so called NBOs. However these QPOs occur at or around the Eddington
mass accretion rate, which is different from the case in
4U\,1323--62. Perhaps our $\sim$1 Hz QPO is related to that reported
by Kommers et al. (1998) in the pulsar 4U\,1626--67, which they
suggest is caused by a structure orbiting the neutron star. However
there is no information on the energy dependence of the 0.048 Hz
QPO. We conclude that so far the $\sim$1 Hz QPO seems to be in a class
of its own.
\newline
\indent Models for the $\sim$1 Hz QPO must explain why the rms
amplitude of the QPO is constant during bursts, dips and
persistent-emission. In models involving a radiation-dominated (inner)
disk, we would expect these properties to be different during a
burst. Therefore, we consider explanations of the QPO in terms of
modulations caused by wave packets of sound waves in the disk (Alpar
\& Yilmaz 1997) and in terms of fluctuations in the electron
scattering optical depth at the critical Eddington mass accretion
rates (Fortner et al. 1989) to be unlikely. Furthermore, assuming
isotropic emission, the 2--25 keV flux in 4U\,1323--62 is too low to
produce these fluctuations in the way proposed by Fortner et
al. (1989).
\newline
\indent Since this QPO is observed in a dipper, we suggest an origin
related to the high inclination at which we observe this system. The
$\sim$1 Hz QPO cannot be formed by a ripple in the same structure that
is also causing the dips, since the fractional rms amplitude of the
QPO should then vary with energy in the same way as the dips. This
seems not to be the case, since during the persistent emission the
spectrum is softer than during the dips, whereas the QPO rms amplitude
increases slightly with photon energy. The fact that the QPO
properties are the same in and outside the dips also makes this model
unlikely. However, quasi-periodic obscuration of the central source by
a structure elsewhere in the accretion flow may cause the QPO. If we
assume that the QPO is caused by structures orbiting the neutron star
with a Keplerian frequency of $\sim$1 Hz, the orbital radius is
$\sim$2$\rm{x}10^8$ cm for a 1.4 $M_{\odot}$ neutron star. A limited
lifetime of the modulating structures, and/or different or shifting
radii can cause the modulation of the X-rays to be quasi-periodic
instead of periodic. Using the fact that no eclipses have been
observed, we derive the height above the plane of the structure
causing the $\sim$1 Hz QPO to be $>5\,10^7$ cm for a lower main
sequence companion and a neutron star of 1.4 $M_{\odot}$. This
scale-height is difficult to explain in view of the Shakura-Sunyaev
thin $\alpha$ disk model. Recently Pringle (1996) discussed a model
where accretion disks in LMXBs are subject to radiation driven
instabilities, causing the accretion disks to warp. If these warps are
responsible for the X-ray modulations as observed, the radius and
structure of the warp may change on viscous time scales, which for
typical parameters is on the order of days to weeks (Frank, King, \&
Raine 1992). Scheduled observations of this source will enable us to
test this hypothesis. 
\newline
\indent Neither inverse Compton scattering nor regular Compton
scattering can cause the observed X-ray modulation on its own. Both
would result in a strong energy dependence of the QPO such as has been
observed for kHz QPOs in, e.g., 4U\,0614+09 (M\'endez et al. 1997),
which is inconsistent with what we observe in 4U\,1323--62. However,
an intermediate-temperature structure, one containing both hot and
cold electrons, or one combining Compton scattering with absorption
could cause the observed modulations.
\newline
\indent A much simpler model is possible in terms of partial covering
of an extended source. Quasi-periodic obscuration of part of this
central source by an opaque medium, for example an orbiting bump on
the disk's surface, can be responsible for modulations with little
energy dependence as observed.

\acknowledgments This work was supported in part by the Netherlands
Foundation for Research in Astronomy (ASTRON) grant 781-76-017. This
research has made use of data obtained through the High Energy
Astrophysics Science Archive Research Center Online Service, provided
by the NASA/Goddard Space Flight Center. We would like to thank Jeroen
Homan for stimulating discussions and the referee for his comments and
suggestions.

\clearpage

\figcaption{\label{figlc} Top: The 2--60 keV, 16-s averaged,
background-subtracted light curve beginning at the start of the first
observation (see Table \ref{obs_log}) showing a dip and a burst. No
dead time corrections have been applied. The statistical uncertainty
is typically 5 c/s. Bottom: The 16 s averaged hardness curve (9.7--16
keV/2--7.9 keV, see text) of the same data. Typical errors in the
hardness are 0.03. The arrows indicate the start and end of the dip in
both the light- and hardness curve.}

\figcaption{\label{fig1Hz} Power density spectrum of the full (2--60 keV)
energy band of the persistent emission of all the observations
combined showing the $\sim$1 Hz peak. The line drawn through the data
points represents the best fit to the data. No dead time corrections
have been applied.}

\figcaption{\label{light_b_1Hz} Light curve showing the QPO during a
burst. Five points correspond to 1.25 s, close to the QPO period.}

\clearpage
\begin{deluxetable}{|c|c|c|c|c|}
\tablecaption{Log of the observations. The average persistent-emission count
rate of each observation is given in column 5.\label{obs_log}}
\startdata
Number & Observation & Date & Time (UTC) & Persistent emission \nl
       &             &      &            & 2--60 keV (c/s/5PCU) \nl
\tableline
1 & 20066-02-01-00 & 25 April 1997 & 22:03--03:56 & 106\nl
2 & 20066-02-01-03 & 26 April 1997 & 05:06--09:41 & 104\nl
3 & 20066-02-01-01 & 26 April 1997 & 22:02--03:55 & 95 \nl
4 & 20066-02-01-04 & 27 April 1997 & 05:08--09:42 & 102\nl
5 & 20066-02-01-02 & 27 April 1997 & 23:12--03:54 & 94 \nl
\enddata
\end{deluxetable}

\begin{deluxetable}{|c|cccc|c|}
\tablecaption{Fractional rms amplitudes (in \%) of the $\sim$1 Hz
QPO. In determining these values the FWHM and frequency were fixed to
0.25 and 0.84 Hz, respectively.\label{fit_table}}
\startdata
Category & & & Energy (keV) & & \nl 
\tableline
 & 2--5.0 & 5.0--6.4 & 6.4--8.6 & 8.6--13.0 & 2--60 \nl
\tableline
Persistent &$8.0\pm0.3$ & $9.9\pm0.5$ & $9.8\pm0.4$ & $10.9\pm0.4$ & 
$9.1\pm0.1$ \nl
Dips       &$8.1\pm0.9$ & $9.2\pm1.1$ &$10.2\pm0.8$ & $9.7\pm0.8$ & 
$8.9\pm0.4$ \nl
Bursts     &$10.3\pm0.9$& $10.2\pm0.9$ & $10.0\pm1.0$ & $10.6\pm1.0$ &
$10.9\pm0.5$ \nl
\enddata
\end{deluxetable}

\clearpage
\begin{figure}[bh]
\centerline{\psfig{figure=lc+hc.ps,width=8cm}}
\end{figure}

\clearpage
\begin{figure}[bh]
\centerline{\psfig{figure=1Hz_pois.ps,width=8cm}}
\end{figure}

\clearpage
\begin{figure}[bh]
\centerline{\psfig{figure=lc_err.ps,width=8cm}}
\end{figure}

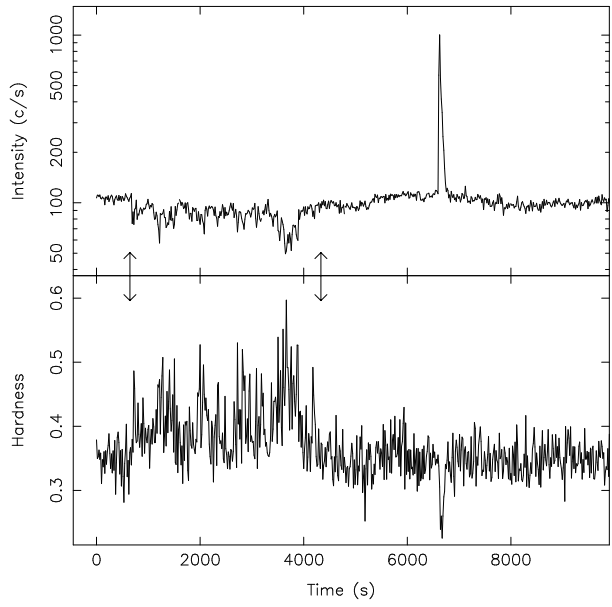
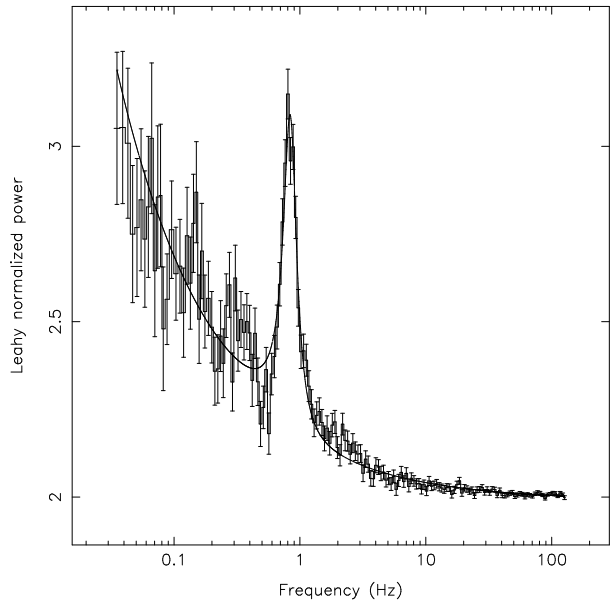
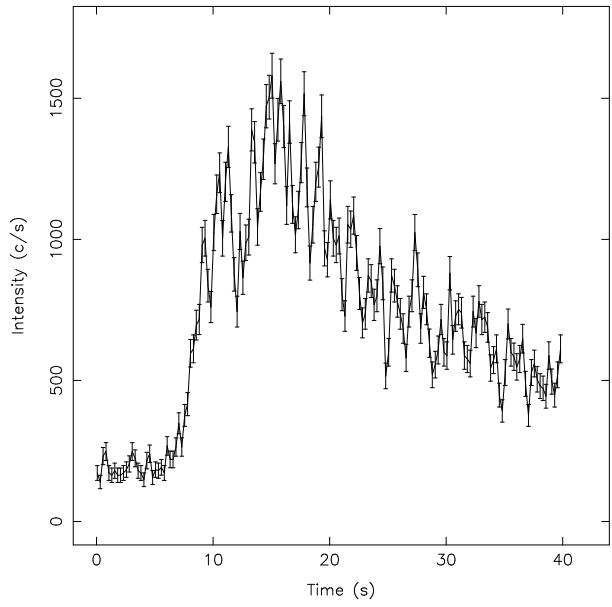
\end{document}